# Nanoscale chiral valley-photon interface through optical spin-orbit coupling

Su-Hyun Gong[1,2], Filippo Alpeggiani[1,2], Beniamino Sciacca[2], Erik C. Garnett[2], L. Kuipers[1,2,*]

[1]Kavli Institute of Nanoscience, Department of Quantum Nanoscience, Delft University of Technology, P.O. Box 5046, 2600 GA Delft, The Netherlands

[2]Center for Nanophotonics, AMOLF, Science Park 104 1098 XG Amsterdam, The Netherlands.

*Correspondence to: L.Kuipers@tudelft.nl

The emergence of two-dimensional transition metal chalcogenide materials has sparked an intense activity in valleytronics since their valley information can be directly encoded and detected by using the spin angular momentum of light. For their practical applications such as on-chip valley logic gates and chip-to-chip valley transport, the encoding and processing of valley pseudospin using light should be extended to an integrated, on-chip nanophotonic system. Here, we successfully demonstrate, at room temperature, the valley-dependent directional coupling of light using a plasmonic nanowire-$WS_2$ layers system. Our calculations show that the local transverse spin angular momentum of the mode of the plasmonic nanowire provides robust optical spin-path locking of up to 91 %. Experimentally we demonstrate that valley pseudospin in $WS_2$ is coupled with optical spin of the same handedness and exhibits a high directional coupling efficiency up to 90 % to the plasmonic guided mode. The result opens up new avenues of controlling, detecting and processing valley and spin information with precise optical control at the nanoscale.

Valleytronics and nanophotonics provide powerful routes to address the heating problem in electronics by offering an alternative to information transport with a charge of electrons (*1-5*). Valley pseudospin provides an additional degree of freedom to encode and process binary information in matter, analogous to the spin degree of freedom. Recently, the study of valleytronics has been stimulated by the emergence of two-dimensional (2D) materials. Particularly, 2D atomically thin transition metal dichalcogenides (TMDs) layers provide a versatile materials platform for both optoelectronics (*6-9*) and valleytronics (*10-14*). They have direct bandgaps consisting of two (energy-degenerate) valleys at the corners of the Brillouin zone (labeled K and K') (*14, 15*). A key requirement as a valleytronics material is the ability to generate and detect the valley polarization (i.e. asymmetric occupation of electrons in two inequivalent valleys). One of the main attractions of TMD materials is that spin and valley information can be optically addressed and detected using the spin angular momentum of light, due to their valley-dependent optical selection rule (*13, 14, 16-18*). Unfortunately, the relatively short lifetimes (< 10 ps) of valley-polarized excitons in TMDs limit logical processes and the spatial transport of valley information (*4*). While valley information is difficult to transport in realistic material systems, particularly at room temperature, photons are ideal information carriers due to the normally small light-matter interaction. This forms the basis of the modern internet and was also recently highlighted in pioneering Bell test/entanglement studies (*19, 20*). Nanoscale coupling of valley pseudospin to photonic degrees of freedom are therefore highly desirable for on-chip integrated valley devices.

The spin angular momentum as a degree of freedom of light at the nanoscale has very recently drawn significant attention since it offers unprecedented ways to influence and exploit light-matter interactions. In highly confined light fields, transverse spin angular momentum, i.e., spin angular momentum that is not parallel but perpendicular to the propagation direction, can naturally appear as a result of the spin-orbit interaction of light (*21-35*). This transverse spin angular momentum provides a robust one-to-one relation between the handedness of optical spin and the propagation direction of a photonic mode−so called spin-momentum locking−due to their time reversal symmetry. As a result, the information of spin angular momentum can be directly transferred to the direction of light and vice versa. Recently, transverse optical spin dependent directional coupling of light with near unity efficiency has been successfully demonstrated experimentally using various photonic structures e.g., optical fibers (*21-23*), metal surfaces (*24-27*), semiconductor waveguides (*29-31*), or microdisks (*32*). Taking this unique property of spin-path locking system one step further, the chiral interaction between local transverse optical spin and a spin- or valley- polarized optoelectronic material, through the proper selection rules, would offer the potential for a wide range of unique applications (*29, 36*), such as chiral-spin entangled quantum networks or single photon optical diodes.

Here, we demonstrate a room-temperature chiral-coupling interface between the transverse optical spin of a plasmonic nanowire mode and the valley pseudospin of tungsten disulfide ($WS_2$), which provides robust valley-polarized directional emission. The highly confined mode of a plasmonic nanowire provides a high degree of local transverse optical spin (~0.91 in our system), and its handedness is locked to the propagation direction of the mode. At any position, a valley-polarized exciton in the TMD layers is preferably coupled to the mode of which the direction is such that its transverse optical spin is the same local handedness of the exciton valley pseudospin. We experimentally prove valley-controlled directional emission at room temperature with a 90% valley-to-photon direction coupling efficiency. The resulting coupling between the photonic path and valley-spin in TMD materials and the metallic nature of the waveguide should pave the way for unique applications in valley-chiral networks, valley-gates, and quantum photonic devices.

Our configuration for a chiral valley-photon interface is schematically depicted in Fig. 1A. It consists of a combination of a few layer TMD and a single silver nanowire. The evanescent fields of the plasmonic guided modes possess transverse spin angular momentum in the plane of the TMD layers (see below for detail). Given valley-dependent optical selection rules, the population of excitons in the two different valleys K and K' can be directly controlled by a circularly polarized excitation laser. Due to the chiral coupling between the circular nature of the emission dipole and the local direction-locked transverse optical spin, emission from the different valleys will couple to plasmonic eigenstates propagating in opposite directions (Fig. 1B). In order to obtain a high degree of valley polarization, we exploit the high spin-valley coupling strength of WS$_2$ (*37, 38*) due to a spin-layer locking effect (*39*). This produces a valley spin splitting for both odd and even layers due to an effective layer decoupling induced by the strong spin-orbit interaction. The degree of valley polarization is denoted by $P_V = (I_K - I_{K'})/(I_K + I_{K'})$, where $I_K$ and $I_{K'}$ indicate the photoluminescence (PL) intensity from K and K', respectively. The number of WS$_2$ layers is identified through their optical contrast in a microscope image and their PL spectrum (*40*) (See SI). The results presented here are based on a WS$_2$ flake consisting of five layers (for results obtained for different layer thicknesses see the SI). Its polarization-resolved spectrum, measured in the absence of the silver nanowire, is shown in Fig. 1C. The measured degree of valley polarizations is found to be 0.7 at room temperature. Figure 1D depicts an optical microscopy image of one of the structures used for investigating valley-path coupling. Figure 1D also presents a spatially resolved image of the collected photoluminescence containing all emission wavelengths. In addition to luminescence from the vicinity of the excitation spot at the center of the waveguide two bright spots at each end of the waveguide demonstrate that emission is coupled to the guided modes of the plasmonic nanowire (*41*). Under local excitation at the center of the silver nanowire, directional emission from the TMD layers is investigated by measuring light scattered at the ends of the wire. The intensity of these bright spots will be investigated as a function of incident polarization and excitation position to demonstrate valley-path coupling.

To quantify the magnitude of the transverse optical spin near a silver nanowire, we performed numerical calculations with a finite difference eigenmode (FDE) solver. Figure 2A depicts the calculated electric field profile of a plasmonic guided mode of the silver nanowire on a glass substrate. The nanowire is modeled with a pentagonal cross-section, which is typical for the chemically grown nanowires (*42*) used in the experiment. The guided plasmonic mode exhibits strong evanescent fields near the interface between the silver nanowire and the glass substrate exactly at the position of WS$_2$ layers. Due to the strong transverse confinement and the plasmonic nature of the modes, a large longitudinal component of the electric field, i.e., a component parallel to the propagation direction (x-direction), is present. The *x* and *y* component of the electric field have comparable amplitudes and are roughly ±90 degrees out of phase: the modes exhibit a large transverse spin angular momentum. The local sign of this phase difference has a one-to-one relation with the propagation direction of light and the position with respect to the mirror plane of the geometry. We calculated the density of the transverse optical spin of the electromagnetic field as a function of the position in the *x-y* plane, which corresponds to the Stokes parameter $S_3 = -2\,\text{Im}(E_x E_y^*)/(E_x^2 + E_y^2)$ (*30*), as shown in Fig 2B. The plasmonic guided modes of infinite length nanowires have a local transverse spin density near unity. The sign of the transverse optical spin on either side of the nanowire is opposite as expected from symmetry considerations. Interestingly, opposite handedness can also be found on the same side of the wire at different distances in the y-direction. This is a consequence of the intricate nature of the plasmonic nanowire's near-field, in the current case primarily caused by the presence of the substrate. For the purpose of this study, it is important to remark that this additional change of handedness occurs relatively far away from the wire ($y$=0.6 μm) where the amplitude of the mode and its influence on the local density of optical states (LDOS) is very small. Clearly, when the propagation direction of the mode is reversed, so are all the signs of the helicity everywhere (not shown). These results are qualitatively summarized in figure 2C that depicts that the handedness of optical spin is determined by a combination of $y$ position (which side of nanowire) and propagation direction (left or right).

To further support our experiments, we simulate the emission from one valley in $WS_2$ in the vicinity of a finite-length plasmonic nanowire with the three-dimensional finite-difference time-domain (FDTD) method. The polarized valley emission is described by the radiation of a circular dipole due to the valley-dependent optical selection rules. Figure 2D depicts the directionality of the emission of a left-handed circularly polarized dipole as a function of its position in the $x$-$y$ plane. The directionality is calculated from the amount of light transmited to both ends of the wire, $D_0 = (T_L - T_R)/(T_L + T_R)$. The first thing to note is that the circular dipole emission is preferentially coupled to a propagating mode with the same handedness of optical spin (compare Fig.2B&C). The second thing to notice is that a high degree of directionality is obtained, which is opposite in sign when exciting on either side of the nanowire. A standing-wave-like pattern is visible along the wire caused by the finite length of the silver nanowire (7 μm) resulting in a small reflection at the end of the wire. Please note that the high value of directionality of 0.91 is still observed despite the presence of a low-amplitude counter-propagating mode which tends to reduce the transverse optical spin and the directionality (*30*). It is important to point out that calculated directionality of a circular dipole, $D_0$, describes the emission with perfect valley polarization. The effective directionality with a finite valley polarization, $P_V$, would be a simple product of $D_0$ and $P_V$. To properly describe the experiment, merely calculating the directionality is not sufficient: we also need to calculate the actual coupling of the emission to the guided modes, Figure 2E depicts the normalized coupling strength, $\kappa_{tot}$, as a function of the position of the circular dipole with respect to the nanowire. $\kappa_{tot}$ is normalized to 1 to facilitate a comparison with the experiment. The high confinement of the modes is reflected in the small size of the region of enhanced LDOS where the emitter couples efficiently. Combining $D_0$ and $\kappa_{tot}$ results in the chiral coupling coefficient $\kappa_{valley-path}$ that describes how well emission from a specific valley couples to a one single direction of the light path: $\kappa_{valley-path} = D_0 \cdot \kappa_{tot}$. The calculated $\kappa_{valley-path}$ is presented in Fig. 2F. It is clear that the experimentally obtained directionality will be dominated by emitting dipoles located close to the nanowire.

For an experimental verification of the chiral valley-photon interface formed by directional coupling of $WS_2$ excitons to plasmonic nanowires, we employ far-field microscopy. Excitons are locally excited at the center of the silver nanowire for different positions in the y-direction with a close-to diffraction-limited focal spot of a 694 nm laser. By exciting with either left- or right-handed circular polarization we generate excitations at either K or K', respectively. For each position an image is recorded of the sample area close to the nanowire. An optical bandpass filter (620-630 nm) is used to predominantly collect the exciton emission. Raw data for a small selection of excitation positions (on either side of the wire and on top) is presented in Figures 3A and 3B for right- and left-handed circular excitation polarization, respectively. Both a full image (top) and a line traces along the center of the wire (bottom) are presented. Like in Figure 1D the large central spot in each image is caused by exciton emission not coupled to the plasmonic nanowire. Coupling of exciton emission to the plasmonic mode of the nanowire is evidenced by the small bright spots on either end of the nanowire, where the light in the propagating mode is scattered to the far field. Comparing the images in Fig. 3A and 3B to Fig. 1D it is clear that the intensity of the spots at the end of the wire have a lower intensity relative to the central excitation spot. This decrease is caused by the fact that Fig. 1D is dominated by the emission from the indirect transition in the near infrared where the propagation losses are significantly less than at 620-630 nm where the direct exciton emission occurs. In the line traces the emission coupled to the nanowire shows up as clear peaks (shoulders) on top of a background corresponding approximately to the cross-section of the excitation spot. Already in this raw data, the valley-controlled directional emission is visible to the "naked eye": for a given combination of handedness and excitation position, an asymmetry is obvious in the intensity of the spots at either end of the nanowire. The asymmetry of emission is reversed when either the excitation spot is moved to the other side of the nanowire or the polarization handedness of the excitation is flipped. If both operations are carried out, the original asymmetry is recovered. These observations prove the concept of valley-dependent directional coupling of light.

In order to quantify the valley-dependent directionality, we determine the PL intensity at the left and right end facets of the scattering spots. After subtracting the background arising mainly from the excitation spot the resulting $I_L$ and $I_R$ reflect the amount of exciton emission that is coupled to the left- and right-propagating plasmonic modes, respectively. The dotted curves in Fig. 3C and 3D show the total guided PL intensity ($I_L+I_R$) as a function of the excitation position. They indicate that the PL is only coupled to the plasmonic modes when the excitation occurs within approximately 500 nm of the nanowire. We then determine the experimental directional coupling efficiency $\kappa_{exp}$: $\kappa_{exp} = (I_L - I_R)/(I_L + I_R)$. Figure 3C and 3D depict $\kappa_{exp}$ as a function of position perpendicular to the wire for left-handed and right-handed circularly polarized excitation, respectively. We observe a clear directional coupling: for right-handed excitation and negative and positive y, the exciton emission couples to plasmonic eigenstates propagating from right to left and left to right, respectively. When the handedness of the excitation is flipped so do the propagation directions of the plasmonic modes to which the emission is coupled. The measurements display exactly the trends in directionality that are observed in the calculations (see Fig. 2F). For symmetric excitation, i.e., excitation on the symmetry axis of the system at the center of the nanowire and far away from the wire $\kappa_{exp}$ is zero, as expected. A small offset in directionality, caused by a small displacement of the excitation spot with respect to the center of the nanowire in combination with propagation losses, was subtracted from the measured result (see the SI). A more quantitative description and comparison with calculations will be presented below. The measurements have been reproduced for a number of plasmonic nanowires and $WS_2$ flakes with different layer thicknesses (see the SI). In all cases, a strong directionality was observed.

In order to confirm that our measured directional emission is indeed caused by chiral interaction between the valley-polarized exciton and the transverse optical spin of the plasmonic nanowire modes, measurements are performed for different excitation polarizations and emission wavelengths. For this a modified experimental set up is used. The direction in the CCD camera image of the luminescence perpendicular to the length of the nanowire is dispersed in wavelength using a grating. This configuration provides position-dependent PL spectrum along the wire from which we can determine $\kappa_{exp}$ for different wavelengths. The results are presented in Figure 4. Figure 4C demonstrates that an equal mix of K and K' excitons, generated by using a linear polarized excitation light, displays no directionality, while for the same nanowire K or K' excitons separately (Figures 4A&B) display valley-controlled directionality. Note that the maximum $\kappa_{exp}$ in Fig. 4A and 4B (~0.17) are lower than the value we observed in Fig. 3 of 0.35. This is caused by the decreased signal to noise ratio, resulting from the fact that the wavelength dispersed image could, by its very nature, not resolve any spatial information along the *y*-direction (see the SI). Figures 4D-F show that the indirect bandgap emission, which does not exhibit valley polarization, does not displays any directionality, regardless of position or excitation polarization. Together these results demonstrate that all experimentally observed directionality is attributable to valley-controlled chiral coupling to plasmonic nanowire modes.

Careful comparison of the experimental and the calculated results reveal two differences. Firstly, the experimental directionality profile is spatially smeared out. Secondly, the magnitude of $\kappa_{exp}$ is less than maximum in the calculations. In part this is caused by obvious experimental limitations, e.g. a finite excitation spot size and background noise: both decrease the experimental maximum of the directionality and the finite excitation spot size also spatially smears out the pattern. However, the reduction in directionality is also caused by the WS$_2$ system itself which exhibits valley-to-valley hopping at time scales commensurate with the exciton lifetime (*43*). This valley-to-valley hopping, captured in the less than zero $P_V$, would actually flip the direction in which the plasmonic modes are launched. These experimental matters are readily taken into account to properly compare our realized chiral valley-photon interface to the ideal simulated interface, using only a single fitting parameter:

$$\kappa_{fit}(y) = \frac{[\kappa_{valley-path}(y) \cdot P_V] \otimes PSF}{[\kappa_{tot} + 2\rho] \otimes PSF},$$

where $\kappa_{valley-path}(y)$ is the calculated chiral valley-path coupling coefficient for *perfect* circularly polarized dipole, $P_V$ is the measured degree of valley polarization (0.7), and $\rho$ is an unpolarized background noise relative to the maximum signal and is the only free fitting parameter. To account for the effect of the finite size of the excitation region, we convolute the calculated signal with the experimentally determined point spread function (PSF) of our laser spot. We obtain excellent agreement between fit and data for $\rho$ =0.006 (gray lines in Fig. 3C&D). With a reduction of directionality through noise as $D_0/(1+2\rho)$ the fitting procedure reveals that the experimental valley-to-path coupling efficiency including background noise is as high as 0.9. The quality of the fit yields a number of important conclusions. Firstly, because the experimentally determined point spread function is sufficient to explain the broadening of the directional coupling efficiency, exciton diffusion does not significantly affects the experiment. Secondly, because the fit uses the calculated $\kappa_{valley-path}$, which implies a perfect circular emission dipole for the valley emission, in combination with the measured finite valley polarization $P_V$ of WS$_2$ at room temperature leads to two more important conclusions. Firstly, any effect of the plasmonic nanowire on the polarization of the excitation focal spot is negligible. Secondly and most importantly, the chiral valley-to-path coupling is only limited by the magnitude of the transverse optical spin of the plasmonic nanowire modes. Thus, we have been able to realize a room temperature interface between the valley pseudospin of WS$_2$ and the propagation direction of nanowire plasmonic modes with a fidelity as high as 0.9.

In conclusion, we have demonstrated the efficient coupling of valley-polarized excitons in $WS_2$ to the transverse optical spin of plasmonic nanowire modes, resulting in valley-photon direction locking with high fidelity. Thanks to the high valley-spin coupling strength of $WS_2$ and high transverse optical spin density in plasmonic modes, the chiral valley (spin)-photon interface is achieved at room temperature and without the necessity of external magnetic fields. This opens avenues for on-chip photon-mediated control of valley information. The one-to-one correspondence of a specific valley state to a corresponding photon path opens new opportunities to gate and transport the valley information using photons with a high speed, solving the challenges of a short lifetime of valley excitons in TMDs. The conducting properties of the plamonic nanowire offer an additional potential connecting the nanoelectronics and nanophotonics. The use of chiral coupling to achieve the interface presents an unprecedented platform for both fundamental studies and wide range of exciting applications of chiral photonics and chiral quantum optics. It is worth noting that this unidirectional, i.e., nonreciprocal, interface can be extended to scalable chiral quantum networks using remarkable spin-valley properties of quantum emitters (*9*) in TMDs.

**References and Notes:**

1. A. Rycerz, J. Tworzydlo, C. Beenakker, Valley filter and valley valve in graphene. *arXiv preprint cond-mat/0608533*, (2006).
2. D. Xiao, W. Yao, Q. Niu, Valley-contrasting physics in graphene: magnetic moment and topological transport. *Physical Review Letters* **99**, 236809 (2007).
3. M. B. Lundeberg, J. A. Folk, Harnessing chirality for valleytronics. *Science* **346**, 422-423 (2014).
4. J. R. Schaibley, H. Yu, G. Clark, P. Rivera, J. S. Ross, K. L. Seyler, W. Yao, X. Xu, Valleytronics in 2D materials. *Nature Reviews Materials* **1**, 16055 (2016).
5. R. Kirchain, L. Kimerling, A roadmap for nanophotonics. *Nat. Photon.* **1**, 303-305 (2007).
6. S. Wu, S. Buckley, J. R. Schaibley, L. Feng, J. Yan, D. G. Mandrus, F. Hatami, W. Yao, J. Vučković, A. Majumdar, Monolayer semiconductor nanocavity lasers with ultralow thresholds. *Nature* **520**, 69-72 (2015).
7. K. F. Mak, J. Shan, Photonics and optoelectronics of 2D semiconductor transition metal dichalcogenides. *Nat. Photon.* **10**, 216-226 (2016).
8. X. Liu, T. Galfsky, Z. Sun, F. Xia, E.-c. Lin, Y.-H. Lee, S. Kéna-Cohen, V. M. Menon, Strong light–matter coupling in two-dimensional atomic crystals. *Nat. Photon.* **9**, 30-34 (2015).
9. A. Srivastava, M. Sidler, A. V. Allain, D. S. Lembke, A. Kis, A. Imamoğlu, Optically active quantum dots in monolayer WSe2. *Nat. Nanotechnol.* **10**, 491-496 (2015).
10. Q. H. Wang, K. Kalantar-Zadeh, A. Kis, J. N. Coleman, M. S. Strano, Electronics and optoelectronics of two-dimensional transition metal dichalcogenides. *Nat. Nanotechnol.* **7**, 699-712 (2012).
11. K. F. Mak, K. L. McGill, J. Park, P. L. McEuen, The valley Hall effect in MoS2 transistors. *Science* **344**, 1489-1492 (2014).
12. Y. Zhang, T. Oka, R. Suzuki, J. Ye, Y. Iwasa, Electrically switchable chiral light-emitting transistor. *Science* **344**, 725-728 (2014).

13. A. M. Jones, H. Yu, N. J. Ghimire, S. Wu, G. Aivazian, J. S. Ross, B. Zhao, J. Yan, D. G. Mandrus, D. Xiao, Optical generation of excitonic valley coherence in monolayer WSe2. *Nat. Nanotechnol.* **8**, 634-638 (2013).
14. X. Xu, W. Yao, D. Xiao, T. F. Heinz, Spin and pseudospins in layered transition metal dichalcogenides. *Nature Physics* **10**, 343-350 (2014).
15. G. Aivazian, Z. Gong, A. M. Jones, R.-L. Chu, J. Yan, D. G. Mandrus, C. Zhang, D. Cobden, W. Yao, X. Xu, Magnetic control of valley pseudospin in monolayer WSe2. *Nature Physics* **11**, 148-152 (2015).
16. H. Zeng, J. Dai, W. Yao, D. Xiao, X. Cui, Valley polarization in MoS2 monolayers by optical pumping. *Nat. Nanotechnol.* **7**, 490-493 (2012).
17. K. F. Mak, K. He, J. Shan, T. F. Heinz, Control of valley polarization in monolayer MoS2 by optical helicity. *Nat. Nanotechnol.* **7**, 494-498 (2012).
18. T. Cao, G. Wang, W. Han, H. Ye, C. Zhu, J. Shi, Q. Niu, P. Tan, E. Wang, B. Liu, Valley-selective circular dichroism of monolayer molybdenum disulphide. *Nature communications* **3**, 887 (2012).
19. B. Hensen, H. Bernien, A. E. Dréau, A. Reiserer, N. Kalb, M. S. Blok, J. Ruitenberg, R. F. Vermeulen, R. N. Schouten, C. Abellán, Loophole-free Bell inequality violation using electron spins separated by 1.3 kilometres. *Nature* **526**, 682-686 (2015).
20. M. Malik, M. Erhard, M. Huber, M. Krenn, R. Fickler, A. Zeilinger, Multi-photon entanglement in high dimensions. *Nat. Photon.* **10**, 248-252 (2016).
21. J. Petersen, J. Volz, A. Rauschenbeutel, Chiral nanophotonic waveguide interface based on spin-orbit interaction of light. *Science* **346**, 67-71 (2014).
22. C. Sayrin, C. Junge, R. Mitsch, B. Albrecht, D. O'Shea, P. Schneeweiss, J. Volz, A. Rauschenbeutel, Nanophotonic optical isolator controlled by the internal state of cold atoms. *Physical Review X* **5**, 041036 (2015).
23. R. Mitsch, C. Sayrin, B. Albrecht, P. Schneeweiss, A. Rauschenbeutel, Quantum state-controlled directional spontaneous emission of photons into a nanophotonic waveguide. *Nature communications* **5**, 5713 (2014).
24. F. J. Rodríguez-Fortuño, G. Marino, P. Ginzburg, D. O'Connor, A. Martínez, G. A. Wurtz, A. V. Zayats, Near-field interference for the unidirectional excitation of electromagnetic guided modes. *Science* **340**, 328-330 (2013).
25. D. O'Connor, P. Ginzburg, F. Rodríguez-Fortuño, G. Wurtz, A. Zayats, Spin–orbit coupling in surface plasmon scattering by nanostructures. *Nature communications* **5**, 5327 (2014).
26. K. Y. Bliokh, D. Smirnova, F. Nori, Quantum spin Hall effect of light. *Science* **348**, 1448-1451 (2015).
27. J. B. Mueller, F. Capasso, Asymmetric surface plasmon polariton emission by a dipole emitter near a metal surface. *Physical Review B* **88**, 121410 (2013).
28. K. Bliokh, F. Rodríguez-Fortuño, F. Nori, A. V. Zayats, Spin-orbit interactions of light. *Nat. Photon.* **9**, 796-808 (2015).
29. A. B. Young, A. Thijssen, D. M. Beggs, P. Androvitsaneas, L. Kuipers, J. G. Rarity, S. Hughes, R. Oulton, Polarization engineering in photonic crystal waveguides for spin-photon entanglers. *Physical review letters* **115**, 153901 (2015).
30. R. Coles, D. Price, J. Dixon, B. Royall, E. Clarke, P. Kok, M. Skolnick, A. Fox, M. Makhonin, Chirality of nanophotonic waveguide with embedded quantum emitter for unidirectional spin transfer. *Nature communications* **7**, 11183 (2016).

31. B. Le Feber, N. Rotenberg, L. Kuipers, Nanophotonic control of circular dipole emission. *Nature communications* **6**, 7695 (2015).
32. F. J. Rodríguez-Fortuño, I. Barber-Sanz, D. Puerto, A. Griol, A. Martínez, Resolving light handedness with an on-chip silicon microdisk. *ACS Photonics* **1**, 762-767 (2014).
33. K. Y. Bliokh, A. Y. Bekshaev, F. Nori, Extraordinary momentum and spin in evanescent waves. *Nature communications* **5**, 3300 (2014).
34. A. Aiello, P. Banzer, M. Neugebauer, G. Leuchs, From transverse angular momentum to photonic wheels. *Nat. Photon.* **9**, 789-795 (2015).
35. T. Van Mechelen, Z. Jacob, Universal spin-momentum locking of evanescent waves. *Optica* **3**, 118-126 (2016).
36. P. Lodahl, S. Mahmoodian, S. Stobbe, A. Rauschenbeutel, P. Schneeweiss, J. Volz, H. Pichler, P. Zoller, Chiral quantum optics. *Nature* **541**, 473-480 (2017).
37. B. Zhu, H. Zeng, J. Dai, Z. Gong, X. Cui, Anomalously robust valley polarization and valley coherence in bilayer $WS_2$. *Proceedings of the National Academy of Sciences* **111**, 11606-11611 (2014).
38. P. K. Nayak, F.-C. Lin, C.-H. Yeh, J.-S. Huang, P.-W. Chiu, Robust room temperature valley polarization in monolayer and bilayer $WS_2$. *Nanoscale* **8**, 6035-6042 (2016).
39. A. M. Jones, H. Yu, J. S. Ross, P. Klement, N. J. Ghimire, J. Yan, D. G. Mandrus, W. Yao, X. Xu, Spin-layer locking effects in optical orientation of exciton spin in bilayer $WSe_2$. *Nature Physics* **10**, 130-134 (2014).
40. W. Zhao, Z. Ghorannevis, L. Chu, M. Toh, C. Kloc, P.-H. Tan, G. Eda, Evolution of electronic structure in atomically thin sheets of $WS_2$ and $WSe_2$. *ACS nano* **7**, 791-797 (2012).
41. A. Akimov, A. Mukherjee, C. Yu, D. Chang, A. Zibrov, P. Hemmer, H. Park, M. Lukin, Generation of single optical plasmons in metallic nanowires coupled to quantum dots. *Nature* **450**, 402 (2007).
42. B. Sciacca, S. A. Mann, F. D. Tichelaar, H. W. Zandbergen, M. A. Van Huis, E. C. Garnett, Solution-phase epitaxial growth of quasi-monocrystalline cuprous oxide on metal nanowires. *Nano letters* **14**, 5891-5898 (2014).
43. R. Schmidt, G. Berghäuser, R. Schneider, M. Selig, P. Tonndorf, E. Malić, A. Knorr, S. Michaelis de Vasconcellos, R. Bratschitsch, Ultrafast coulomb-induced intervalley coupling in atomically thin $WS_2$. *Nano letters* **16**, 2945-2950 (2016).

**Acknowledgments:** We acknowledge funding from ERC Advanced, Investigator Grant No. 340438-CONSTANS. This work is part of the research program of The Netherlands Organization for Scientific Research (NWO).

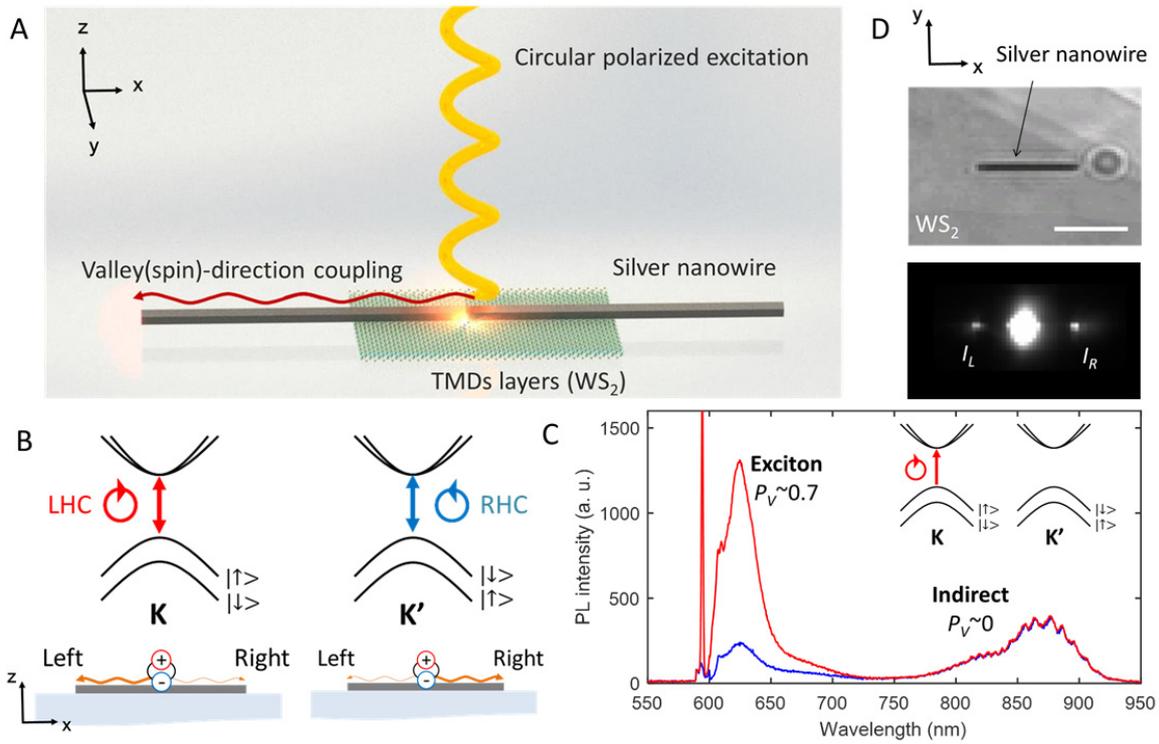

**Fig. 1**. Valley-controlled directional coupling of light (A) A conceptual illustration of directional emission of a valley-polarized exciton in $WS_2$. The valley pseudospin and photon-path are coupled by means of spin-orbit coupling of light. (B) A schematic sketch of the band diagram of $WS_2$ and its optical selection rules depending on the valley index. The two opposite-handed circularly polazired emissions from each valley would couple preferentially to mode propagating in opposite directions. (C) Polarization-resolved emission spectrum of $WS_2$, measured without a silver nanowire, at room temperature excited by a left-handed circularly polarized laser beam at 594 nm. Red and blue spectra indicate left and right handed circularly polarized emission, respectively. (D) Image of a fabricated $WS_2$-silver nanowire coupled system (top). $WS_2$ emission coupled to the plasmonic waveguide mode and scattered at the ends of the nanowire (bottom). The scale bar indicates 5 μm.

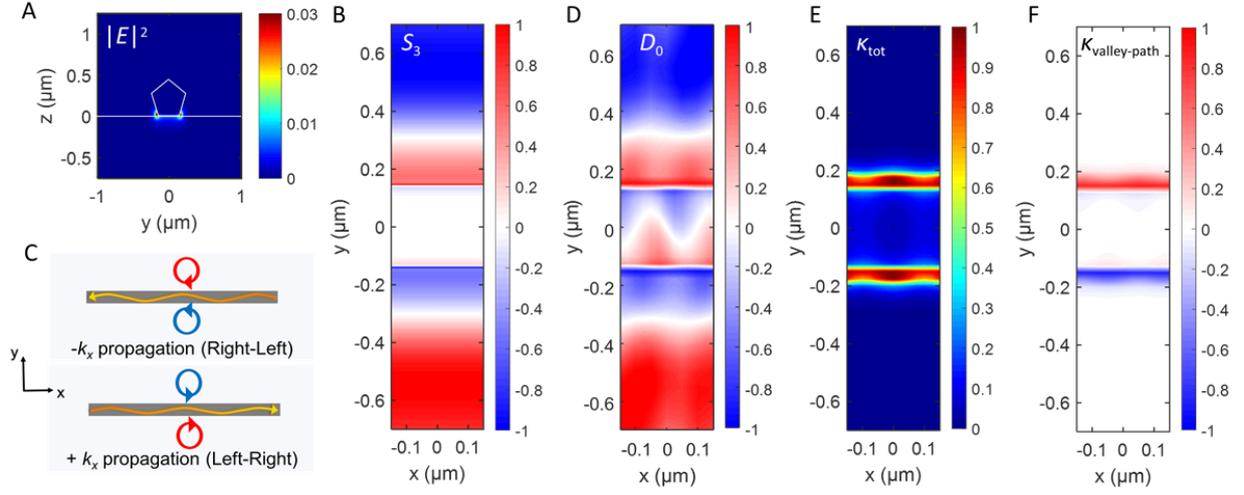

**Fig. 2** Numerical modeling of the transverse spin angular momentum of light (A) Cross-sectional electric field distribution of the plasmonic guided mode. (B) Density of transverse optical spin, i.e. Stokes parameters $S_3$, of the in-plane (*x-y* plane) electric field component. (C) Illustration of the distribution of handedness of elliptical polarization, optical spin, which is depending on both the position and propagation direction. (D) Directionality of circularly polarized dipole emission, $D_0(r)$, as a function of its position. (E) Normalized coupling strength of circularly polarized dipole source to the plasmonic guided modes, $\kappa_{tot}(r)$ (F) Chiral coupling coefficient indicating how efficient circular dipole in one valley couples to a one direction of the guided mode, $\kappa_{valley-path}(r) = D_0(r) \cdot \kappa_{tot}(r)$.

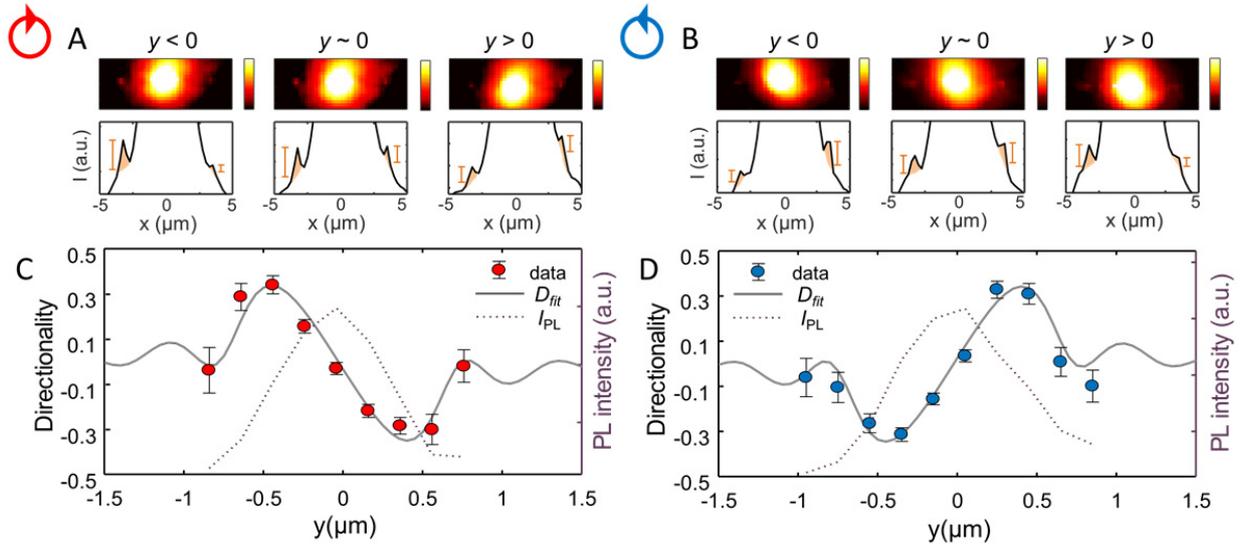

**Fig. 3** Experimental demonstration of valley-controlled directional emission. (A and B) Fluorescence images of the emission of valley-polarized excitons (log scale; top) and line cuts of the intensity profiles (linear scale; bottom) along the silver nanowire under left- (A) and right- (B) handed circularly polarized-excitation. The intensities at the excitation regions were deliberately saturated for clarity of the images. The length of the silver nanowire is ~7 μm. (C and D) Measured directional coupling efficiency, $(I_L - I_R)/(I_L + I_R)$, of the guided emission as a function of the position of the excitation laser with the left- (C) and right- (D) handed circular polarization. Gray lines represent fitting results using the calculated directional coupling efficiency. Purple dotted lines correspond to the total PL intensity measured from the ends of the nanowire as a function of the excitation position.

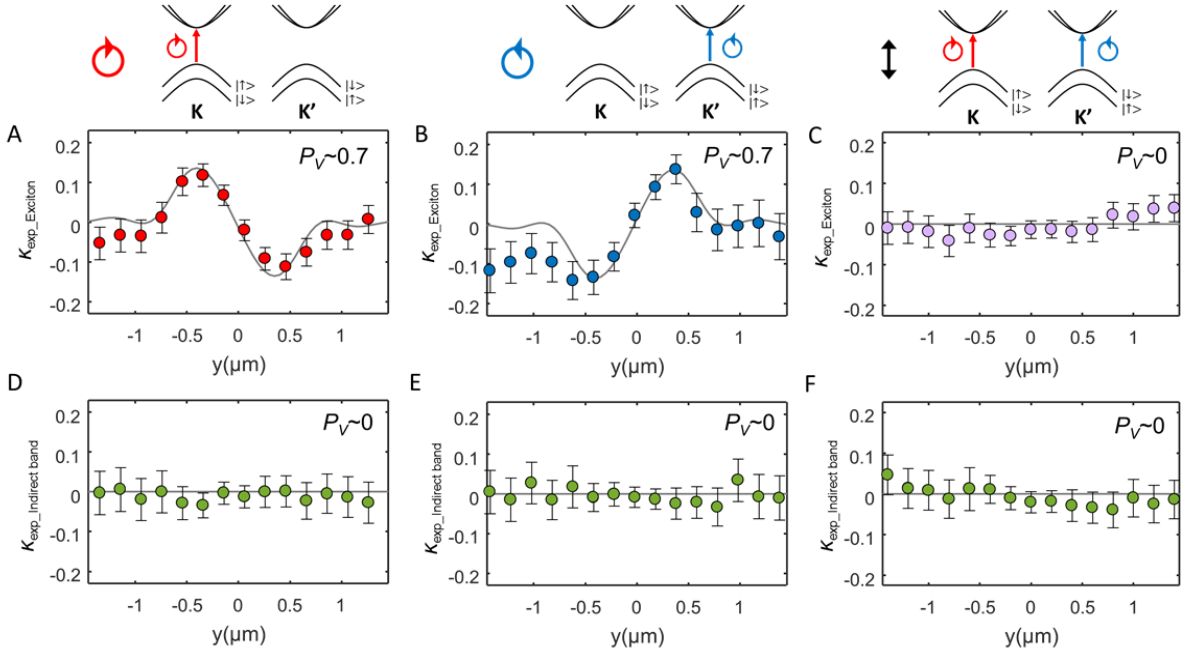

**Fig. 4** Direct observation of directional emission induced by valley-path coupling of the valley-polarized exciton. The emission from the valley-polarized excitons shows directional emission which depends on both the position of excitation and the handedness of the circularly polarized laser (A, B), while the transition of the indirect bandgap has a non-directionality for both handednesses of the circularly polarized laser (D, E). A linearly polarized laser excitation leads to zero valley polarization so that both the exciton- (C) and indirect band- (F) transition exhibit no directionality. $P_V$ indicates the degree of valley polarization, $P_V = (I_K - I_{K'})/(I_K + I_{K'})$.

# Supplementary Materials for

## Nanoscale chiral valley-photon interface through optical spin-orbit coupling

Su-Hyun Gong[1,2], Filippo Alpeggiani[1,2], Beniamino Sciacca[2], Erik C. Garnett[2], L. Kuipers[1,2,*]

correspondence to: L.Kuipers@tudelft.nl

**This PDF file includes:**

Materials and Methods
Supplementary Text
Figs. S1 to S7

**Supplementary Materials:**

**Materials and Methods**

Sample preparation

Few layers of $WS_2$ are exfoliated from commercially obtained (2D semiconductor) bulk $WS_2$ and transferred on a cover glass substrate using a magic tape. Chemically synthesized silver nanowires (41) are positioned on the $WS_2$ layers using a stamping method with a polydimethylsiloxane (PDMS) stamp. The number of $WS_2$ layers is identified through their optical contrast in a microscope image and their PL spectrum (See Fig. S4).

Numerical modeling

Numerical modeling is conducted with both a finite difference eigenmode (FDE) solver and a three-dimensional finite-difference time-domain (FDTD) method, with commercial software (Lumerical). The degrees of circular polarization of the plasmonic field as a function of the position are calculated using $x$- and $y$- electric field components, which are parallel to the layer of $WS_2$. Directional emission from a circular dipole source is determined by calculating transmitted powers at the ends of the nanowire. The circular dipole source is modeled by putting two orthogonal linear polarized-dipole sources with a phase difference of the 90 degrees.

Experimental setup

The photoluminescence of $WS_2$ is induced at room temperature using a semiconductor diode laser with a wavelength of 594 nm. An oil-immersion micro-objective lens with a numerical aperture of 1.45 is used to focus the excitation laser and to collect the emission from $WS_2$. The spectrum of $WS_2$ emission is measured by cooled CCD camera attached to a monochromator. All optical images are measured with an optical long-pass filter (> 600 nm) to block the excitation laser. The position of the sample is controlled with a motorized micrometer stage with a minimum step size of 200 nm. The polarization of the excitation laser and PL are controlled by a half and a quarter waveplate while a fixed polarizer is used to avoid the errors caused by the polarization-dependent collection efficiency of the optical setup. Photoluminescence image of the valley-polarized exciton is obtained by using a spectral filter with the transmission wavelength range of 620-630 nm. For the study of wavelength-resolved directional emission, the y-axis of the image, perpendicular axis to the wire, of the sample is dispersed using a grating (150 g/mm, 300 nm braze wavelength) inside the monochromator (Fig. S1).

**Supplementary Text**

1. The relation between the density of the transverse optical spin of the guided mode and the directionality of circular dipole radiation

The directional emission of the circular dipole can be described by the sum of two coupling efficiencies for opposite handedness of local chirality of electromagnetic fields. The left-handed circular dipole, which is placed on the *x-y* plane, is defined as $\hat{d} = 1/\sqrt{2} \cdot (\hat{x} + i\hat{y})$. For each position in the *x-y* plane of the fields, the plasmonic guided modes can be decomposed into two elliptical polarizations associated plasmonic modes propagating opposite directions, $\widehat{E}_{\pm k} = 1/\sqrt{A_x^2 + A_y^2} \cdot (A_x \hat{x} + A_y e^{\pm i\phi} \hat{y})$. The emission rates of the circular dipole source to the two opposite handedness of elliptical modes are given by,

$$P_{\pm k} \propto \text{Re} \left| \hat{d}^* \cdot \widehat{E}_{\pm k} \right|^2 = \text{Re} \left| \frac{1}{\sqrt{2}} (\hat{x} + i\hat{y})^* \cdot \frac{1}{\sqrt{A_x^2 + A_y^2}} (A_x \hat{x} + A_y e^{\pm i\phi} \hat{y}) \right|^2$$

Then, we define directionality as follow $\frac{P_{-k} - P_{+k}}{P_{-k} + P_{+k}} = \frac{2 A_x A_y \sin\phi}{A_x^2 + A_y^2}$. One can notice from this results that directionality of circular dipole emission has the same value as the local density of the transverse optical spin of the electric fields on the *x-y* plane.

2. Effect of a finite length of the silver nanowire

For an experimental demonstration of valley-dependent directional emission, we utilize a nanowire of finite length. The finite length of the silver nanowire not only results in scattered light originating from excitons coupled to plasmonic modes propagating in one direction, which we use in the measurements to determine the directionality. It also leads to modes with an opposite propagation direction due to mode reflection at the wire end. This influences both the density of the transverse optical spin of a plasmonic mode and the effective, experimentally measured directionality of a circular dipole source.

  a. The density of the transverse optical spin

Figure S2 shows calculated the density of the transverse optical spin for both infinite and finite length of nanowires. In the case of the nanowire with a finite length (~7 μm), the electric fields exhibit an interference pattern due to the reflected guided light. Consequently, both intensity variations and variations in the phase difference between $E_x$ and $E_y$ are introduced as compared to the infinite nanowire. The phase difference of $E_x$ and $E_y$ for infinite silver nanowire is roughly 90 degrees while it lies in the range of 72~78 degrees, depending on the position, for the finite silver nanowire. As a result of the variation of intensity and phase difference of $E_x$ and $E_y$, the density of the transverse optical spin for a finite length of nanowire will have lower magnitude than that of the infinite nanowire. (Note that the period of the interference pattern is much smaller than the size of excitation spot so that in the experiment we only observe an average effect.) As a result of the finite length the maximally attainable directionality reduces from approximately 1 for an infinite nanowire to 0.91 for a 7 μm long nanowire.

b. Effect of reflections on directionality measurements

As we discuss in S1, the density of the transverse optical spin has a direct relation with the directional emission of a circular dipole source. In the case of a finite silver nanowire, however, the measured directionality could seem less than actual directionality of the emission. This measurements effect is caused by reflections at the nanowire ends. As we determine the experimental directionality from the light scattered from the nanowire end, any guided light reflected at that end and propagating to the other end has the potential of reducing the perceived directionality. To determine the magnitude of this effect on the measured directionality, the time-dependent electric field is calculated with the length of 7 μm as shown in Fig. S3. It turns out that the intensity arriving at opposite end of the nanowire after one reflection is less than 3 % compared to the original directional emission. With increasing nanowire length, the magnitude of this effect will be reduced through propagation losses as is shown in Fig. S4.

3. The effect of size of the excitation laser on the WS$_2$

As shown in Fig. 2E, the directional emission of the valley-polarized exciton is depending on the position with respect to the silver nanowire. Since the size of our excitation focus is relatively large compared the spatial extent of the plasmonic guided modes, we need to consider the total directional emission from an extended spatial area. The total guided emission to the left and right end of the silver nanowire from the extended excitation area can be written as $I_{Total\_L} = \iint dxdy [I_L(x,y) + \alpha]$ and $I_{Total\_R} = \iint dxdy [I_R(x,y) + \alpha]$, respectively, where $\alpha$ is non-directional background noise without directionality. Then, the measured directionality of the exciton emission is as follow.

$$\kappa_{\exp} = \frac{\iint dxdy [I_L(x,y) - I_R(x,y)]}{\iint dxdy [I_L(x,y) + I_R(x,y) + 2\alpha]}$$

We can estimate the effective directionality if the background signal is absent using $[I_L(x,y) - I_R(x,y)]/[I_L(x,y) + I_R(x,y)] = D_0(x,y) \cdot P_V$, where $D_0(x,y)$ is calculated directionality of single circular dipole and $P_V$ is the degree of valley polarization. Then,

$$\kappa_{\exp} = \frac{\iint dxdy [(I_L(x,y) + I_R(x,y)) \cdot D_0(x,y) \cdot P_V]}{\iint dxdy [I_L(x,y) + I_R(x,y) + 2\alpha]} = \frac{\iint dxdy [I_{Total}(x,y) \cdot D_0(x,y) \cdot P_V]}{\iint dxdy [I_{Total}(x,y) + 2\alpha]}$$

where $I_{Total}(x,y) = I_L(x,y) + I_R(x,y)$. $I_{Total}(x,y)$ is governed by both the coupling strength to the plasmonic modes ($\kappa_{tot}$) and the local power of excitation laser $I_{ex}(x,y)$, i.e., $I_{Total}(x,y) \propto k_{tot}(x,y) \cdot I_{ex}(x,y)$. The effect of spatial distribution of excitation power, $I_{ex}(x,y)$, is accounted for through the point spread function (*PSF*) of our focused laser spot. As a result, we can fit the measurements of $\kappa_{\exp}$ to obtain the directional coupling efficiency as a function of the position (y) of the laser as follows

$$\kappa_{fit}(y) = \frac{[\kappa_{tot}(y) \cdot D_0(y) \cdot P_V] \otimes PSF}{[\kappa_{tot}(y) + 2\rho] \otimes PSF} = \frac{[\kappa_{valley-path}(y) \cdot P_V] \otimes PSF}{[\kappa_{tot}(y) + 2\rho] \otimes PSF}$$

where κ$_{valley-path}$ is the calculated chiral coupling coefficient that describes how well emission from a specific valley couples to a one single direction of the light path, i.e., $\kappa_{valley-path}(y) = \kappa_{tot}(y) \cdot D_0(y)$, and $\rho$ is the unpolarized background noise normalized by maximum intensity of the signal, i.e., $\rho = \alpha/I_{tot}$.

When the excitation spot is located not exactly in the middle along the nanowire, i.e., $x \neq 0$, the excited leftward and rightward propagating modes will undergo different propagation losses as they reach the ends of the wire. This cause a y-independent shift of $\kappa_{exp}$. Fig. S5A illustrates an expected directional coupling efficiency with shifted offsets of $\kappa_{fit}(y)$ depending on the x position. The raw data of Fig. 3C shows a good agreement with the fitting results with x=-200 nm.

4. Measured directionality depending on the number of WS$_2$ layers.

Due to their high spin-valley coupling constant, WS$_2$ can emit the light with a high degree of circular polarization even with the even number of layers (*37-39*). Figure S6 shows the measured degree of circular polarization as a function of the number of WS$_2$ layers. It reveals a high degree of circular polarization of ~0.6 for more than three layers of WS$_2$, while monolayer and bilayer have a degree of circular polarization of ~0.08 and ~0.28, respectively.

We also have measured the directional emission to plasmonic guided modes with three different thickness of WS$_2$ layers, as shown in Fig. S7. The directionality is conducted with a CCD camera attached to a monochromater to investigate wavelength dependent directionality. For all samples, indirect bandgap transition shows non-directionality as a function of the position of laser (data is not shown). In the case of the monolayer with valley polarization of 0.08, the directionality of exciton emission could not be resolved by our detection system. The measured directionality is not perfectly linear as a function of the valley polarization since directionality is determined by not only valley polarization but also signal to background ratio, which is related to the photoluminescence intensity of WS$_2$ layers, the propagation loss of plasmonic modes, and the length of the silver nanowire etc.

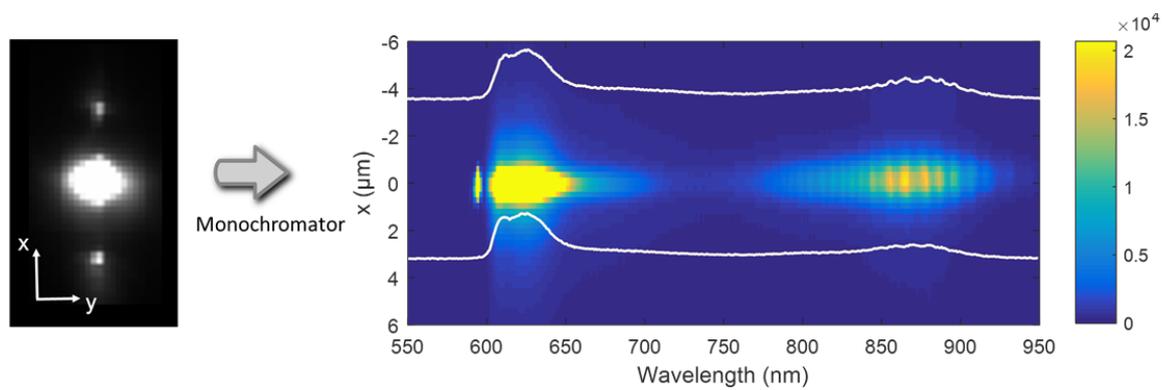

**Fig. S1** One example of measured x position dependent spectrum array obtained using a 2D CCD camera attached to a monochromator. By extracting spectrum distribution at the ends of the silver nanowire, the wavelength-resolved directional emission is investigated.

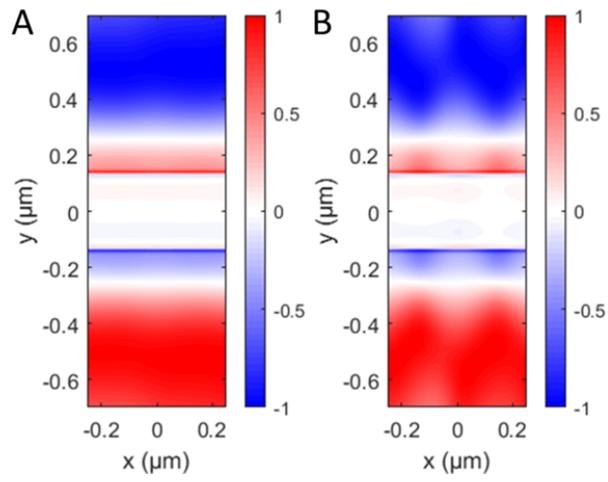

**Fig. S2** The density of the transverse optical spin near a silver nanowire calculated using FDTD methods without (A) and with (B) the effect of reflected light at the end of the silver nanowire.

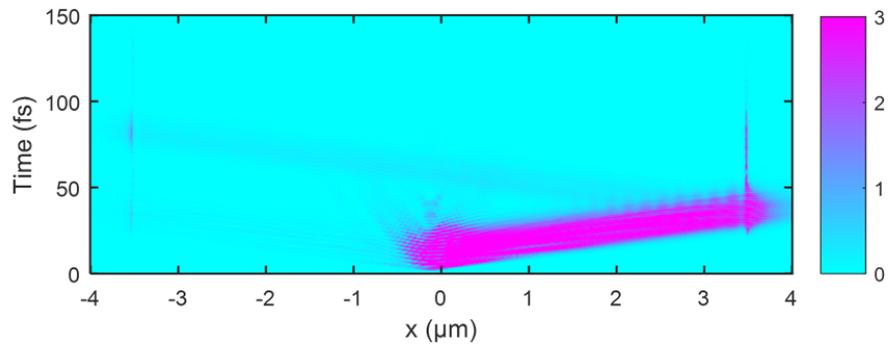

**Fig. S3** The calculated time-dependent electric field distribution along the wire. The circular dipole is placed at the position (x=-100 nm y=150 nm) where it has maximum directional emission, near 1, in the Fig. 2E.

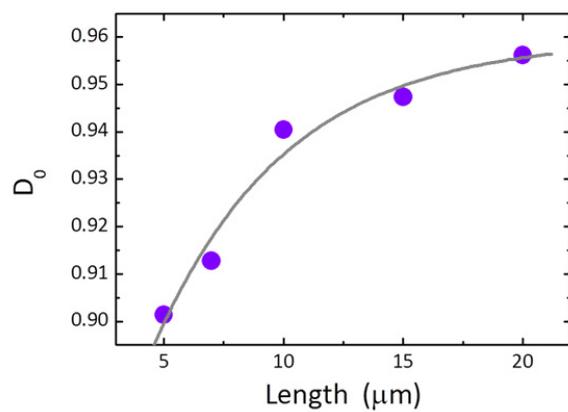

**Fig. S4** The calculated maximum directionality as a function of the nanowire length.

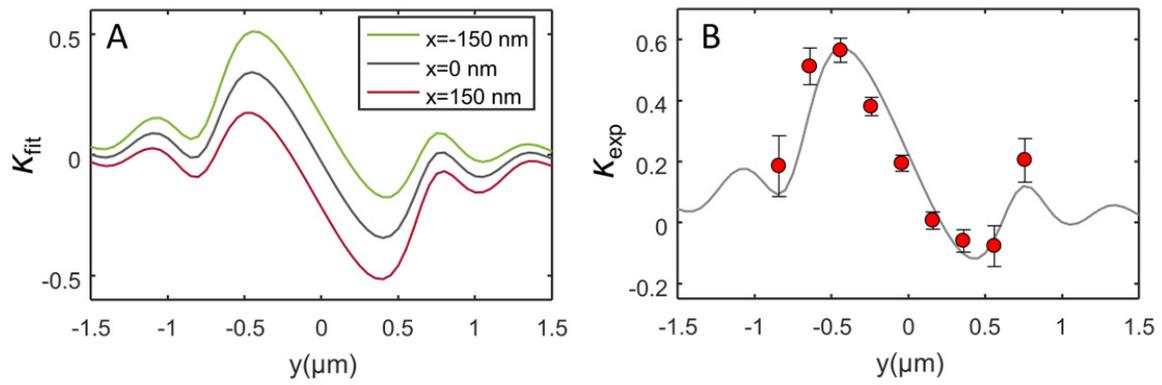

**Fig. S5** The effect of position of excitation along the nanowire on directional coupling efficiency of valley polarized exciton. (A) Calculated directional coupling efficiency depending on the position of a circular dipole. (B) The fitting results with a raw data of Fig. 3A. The excitation spot of laser is 200 nm off from the center of the wire.

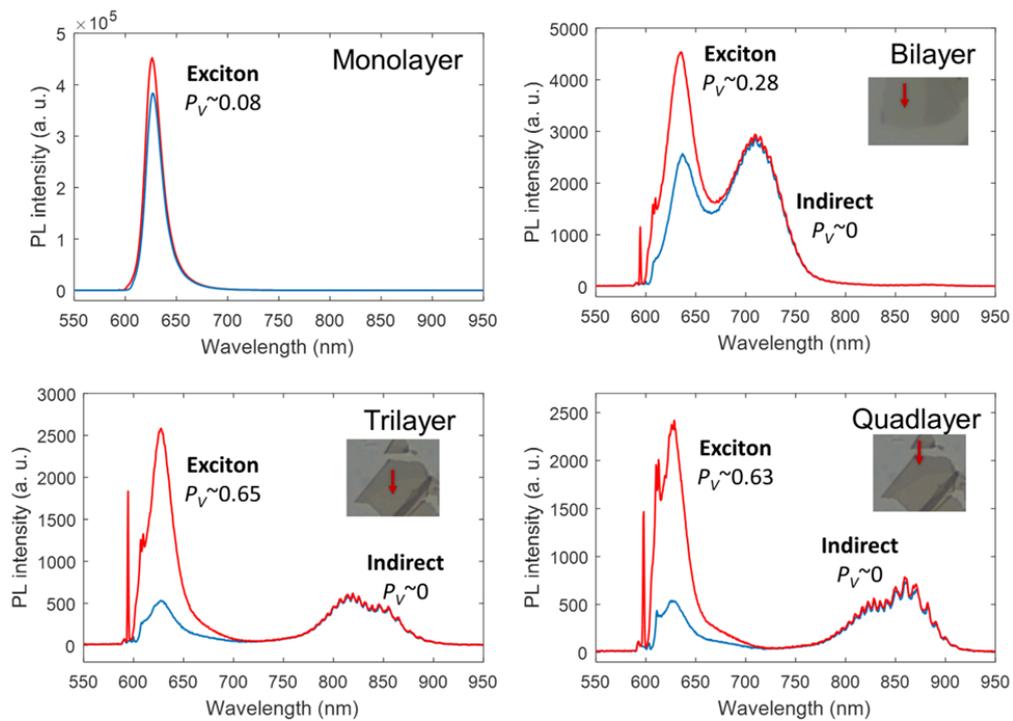

**Fig. S6** The polarization dependent spectrum from difference number of $WS_2$ layers. The number of layers is determined by measuring the energy of indirect transition of $WS_2$ and optical contrast of microscope image.

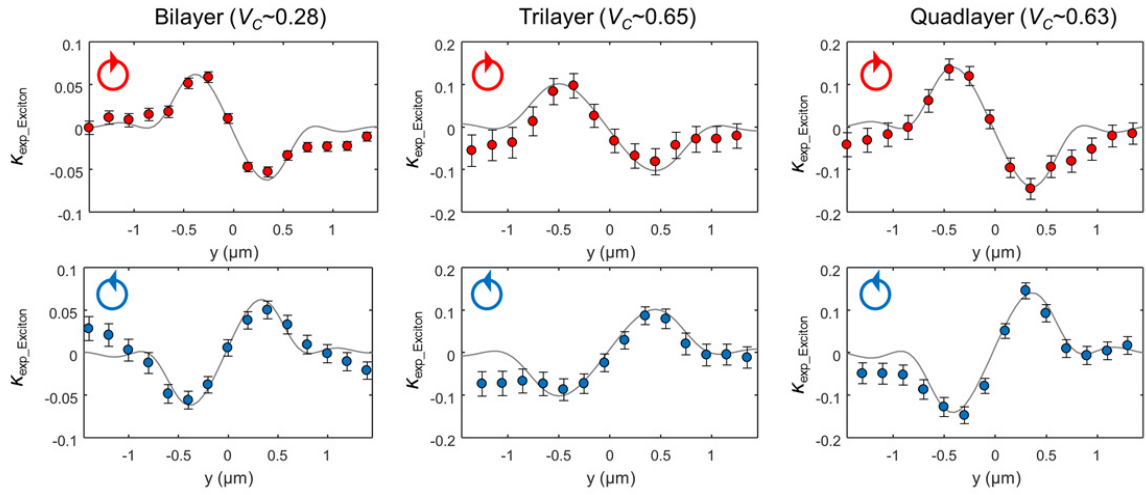

**Fig. S6** Measured directional coupling efficiency of exciton emission depending on the number of $WS_2$ layers, which is estimated from dispersed images using grating. The gray lines indicate the fitting results with a theoretical calculation.